\documentclass{article}

\usepackage[]{graphicx}

\newcommand{\be}{\begin{equation}}
\newcommand{\ee}{\end{equation}}

\begin{document}


\title{Surface brightness in plasma-redshift cosmology}         
\author{Ari Brynjolfsson \footnote{Corresponding author: aribrynjolfsson@comcast.net}}

\date{\centering{Applied Radiation Industries, 7 Bridle Path, Wayland, MA 01778, USA}}          

\maketitle

\begin{abstract}
In 2001 Lori M. Lubin and Allan Sandage, using big-bang cosmology for interpreting the data, found the surface brightness of galaxies to be inversely proportional to about the third power of (1+z), while the contemporary big-bang cosmology predicts that the surface brightness is inversely proportional to the fourth power of (1+z).  In contrast, these surface brightness observations are in agreement with the predictions of the plasma-redshift cosmology.  Lubin and Sandage (2001) and Barden et al. (2005), who surmised the big-bang expansion, interpreted the observations to indicate that the diameters of galaxies are inversely proportional to (1+z).  In contrast, when assuming plasma-redshift cosmology, the diameters of galaxies are observed to be constant independent of redshift and any expansion.  Lubin and Sandage (2001) and Barden et al. (2005), when using big-bang cosmology, observed the average absolute magnitude of galaxies to decrease with redshift; while in plasma redshift cosmology it is a constant.  Lubin and Sandage and Barden et al.~suggested that a coherent evolution could explain the discrepancy between the observed relations and those predicted in the big-bang cosmology.  We have failed to find support for this explanation.  We consider the observed relations between the redshift and the surface-brightness, the galaxy diameter, and the absolute magnitude to be robust confirmations of plasma-redshift cosmology.

\end{abstract}




\makeatletter	   
\renewcommand{\ps@plain}{
     \renewcommand{\@oddhead}{\textit{Ari Brynjolfsson: Surface brightness in plasma-redshift cosmology}\hfil\textrm{\thepage}}%
     \renewcommand{\@evenhead}{\@oddhead}
     \renewcommand{\@oddfoot}{}
     \renewcommand{\@evenfoot}{\@oddfoot}}
\makeatother     

\pagestyle{plain}


\section{Introduction}

Big-bang cosmology is the commonly accepted cosmology, see for example Peebles' monograph on the subject [1].  Most observation, such as the magnitude-redshift relation for supernovae type Ia (SNe\,Ia) [2-9], are explained on the basis of the big-bang cosmology.  Cosmic time dilation factor equal to the expansion factor $(1+z)$ is an integral part of the big-bang expansion hypothesis.

\indent  For determining the comoving distances (see [1,\,10,\,11,\,12]), the big-bang cosmology uses besides the Hubble constant $H_0 {\rm{,}}$ three parameters, $\Omega_k ,\, \Omega_m \,{\rm{and}}\,\Omega_{\Lambda} {\rm{.}}~$ These four parameters are then adjusted to fit each observation. 
If the big-bang hypothesis is correct, we expect that different observations lead to the same value of these parameters.

\indent  Plasma-redshift of photons in hot sparse plasma is a newly discovered cross section, which explains the cosmological observations through means of conventional physics.  Plasma redshift is not a hypothesis or a conjecture invented for explaining something.  Instead, it follows from or is a consequence of conventional axioms of physics without any new assumptions.  Our calculations of the photons interactions with the hot sparse plasma are only more exact than those usually found in the literature [13]; see in particular sections 1 through 4 and Appendix A. 

\indent  The energy lost by the photons in the plasma redshift is absorbed in the plasma.  The steep temperature rise from about 6,000 K in the solar photosphere to more than 2 million K in its corona is mainly caused by the plasma redshift heating; see section 5 of [13].  Like in the Sun, the plasma-redshift heating is important in the coronas of stars, galaxies, and quasars.  This redshift heating is the principal cause of the heating of sparse intergalactic plasma to an average electron temperature of about $ T_e \approx 2.7\cdot 10^6$ K.  Without the plasma redshift these high temperatures could not be explained.  Because big-bang cosmologists could not find any explanation for the observed cosmological redshifts, they surmised that these redshifts were caused by Doppler shifts due to a surmised expansion of the universe.  They could not explain the intrinsic redshifts of stars, galaxies, and quasars.  They explained all such redshifts as being caused by Doppler shifts, in spite of much evidences showing that these objects had intrinsic redshifts. 

\indent  Plasma redshift is not significant in atoms or cold plasmas with densities used in conventional laboratory experiments.  This is most likely the main reason for why it was not discovered before.

\indent  Plasma-redshift cosmology explains the observed magnitude-redshift relation [7]\,and\,[9] and the entire cosmological redshift [12\,-\,17].  There is no need for the big-bang expansion, the dark energy, or dark matter.  Plasma redshift explains the cosmic microwave background (CMB) and the cosmic X-ray background; see sections 5.10 and 5.11, and Appendix C of [13]. 

\indent  In section 2, we summarize the methods used and the findings by Lubin and Sandage [18\,-\,21].   
Their analyses of the experiments indicated that the surface brightness of galaxies is proportional to about $(1+z)^{-3}$ and not $(1+z)^{-4}$ as required by the big-bang cosmology.  In section 3, we show how to translate the specific big-bang cosmology used by Lubin and Sandage to any other cosmology for easy comparison with observations.  From this it can clearly be seen that the observations are consistent with plasma-redshift cosmology.  In section 4, we discuss why the suggested evolution models, which have been proposed to explain the discrepancy between observations and the big-bang cosmology, are false. In section 5, we summarize the main conclusions.  For facilitating the explanations, we review briefly in Appendix A the distance-redshift relations in the different cosmologies.


\section{The Tolman's brightness tests by Lubin and Sandage}

Tolman [22]\,and\,[23] showed that in the big-bang cosmology, the surface brightness (the light intensity per unit area), is inversely proportional to the expansion factor in the fourth power, $i_{\langle SB \rangle}\propto (1+z)^{-4} {\rm{;}}$ see Eq.\,(A14) of the Appendix A. 

\indent  For quantifying the surface brightness, Lubin and Sandage used the Petrosian [24] definition of the parameter $\eta$ as a function of the radius $r$ and the luminosity $L(r)$ at $r$ of the object; and they used the methods developed by Djorgovski \& Spinrad [25], Sandage and Perlmutter [26], and Sandage [27].   The $\eta$ function is given by
\be
\eta = 2.5\, \log \, [2 d(\log \,r)/d\{\log\,L(r)\}]\, {\rm{.}}
\ee
\noindent  This function is helpful in defining the area of interest.  By analyzing several nearby galaxies Lubin and Sandage were able to determine $\eta$ as a function of the surface brightness at $r {\rm{.}}~$ Usually, the radius of a galaxy and its brightness are well defined at $\eta = 1.7$ and $\eta = 2.0 {\rm{,}}~$ which therefore are the $\eta$-values usually used by Lubin and Sandage.  The determination of the brightness variation with $z$ is a difficult task, because the observation require a narrow angle and elaborate corrections for absorption and spectral variations.  Lubin and Sandage, however, are highly experienced in estimating these corrections. 

\indent  In addition to their analyses of many low redshift galaxies they also analyzed many galaxies in each of 3 high-redshift clusters: Cl~1324+3011 at a redshift of $z=0.7565{\rm{,}}$ Cl~1604+4304 at $z=0.8967 {\rm{,}}$ and Cl~1604+4321 at $z=0.9243{\rm{.}}~$ They use the big-bang model and Eq.\,(A14) for surface brightness as a reference.  In the denominator on the right side of Eq.\,(A14) they replaced the factor $(1+z)^{4} {\rm{}}$ by $(1+z)^n$ and determined $n$ experimentally. 

\indent  Sandage and Lubin [18] pointed out that the surface brightness $\langle SB \rangle_{\eta}$ in magnitude units is
\be
\langle SB \rangle_{\eta} = 2.5 \, \log (\pi r_{\eta}^2) \quad+\quad m_{\eta} {\rm{,}}    
\ee
where $ m_{\eta}$ is the magnitude of the light within $r_{\eta} ~{\rm{arcsec .}}$  The beauty of this equation is that all quantities are directly observed and are independent of all cosmologies. 

The theoretical interpretation of the experiments depends on the distances in the different cosmologies; see Eqs.\,(A1) to (A6).  For the initial evaluation, we use with Lubin and Sandage Eq.\,(A14) for surface brightness, and the distances $D_{bb}$ estimated by Mattig's equation for the parameter $q_0 = 0.5 {\rm{,}}$ Eq.\,(A6), and a Hubble constant $H_0 = 50 ~{\rm{km\,s}}^{-1}\,{\rm{Mpc}}^{-1}{\rm{.}}~$  

\indent  A different formulation, often used, characterizes the cosmology with the parameters $\Omega_{\kappa},\,\Omega_m,\, \Omega_{\Lambda}{\rm{.}}~ $  The distances corresponding to the Mattig's equation with $q_0=0.5{\rm{,}}$ Eq.\,(A6), are then given by Eq.\,(A4) and $(\Omega_{\kappa},\,\Omega_m,\, \Omega_{\Lambda}) =(0.0,\,1.0,\, 0.0){\rm{.}}~$ In Fig.\,(1), (2), and (3) this model corresponds to the big-bang parameters $(\Omega_m,\, \Omega_{\Lambda}) =(1.0,\, 0.0){\rm{.}}~$ This reference model, which was used by Lubin and Sandage [18\,-\,21], requires adjustments for accurate evaluation when the results of experiments are applied to other cosmologies, including the plasma-redshift cosmology.  In section 3, we discuss these necessary adjustments for other models.

\indent  From the observations of the galaxies in the R band and I band, respectively, they derived the change in brightness magnitude, $\Delta \langle SB \rangle {\rm{,}}$ to be (see their Eq.\, (7) of [21])
\be
\Delta \langle SB \rangle_R = 2.5\, \log\,(1+z)^{2.59 \pm 0.17}\quad {\rm{and}}\quad \Delta \langle SB \rangle_I = 2.5\, \log\,(1+z)^{3.37 \pm 0.13} \,{\rm{,}}
\ee 	  
\noindent  The average of the R-band and the I-band is $(2.59+3.37)/2 = 2.98 $ or 
\be
\Delta \langle SB \rangle = 2.5\, \log\,(1+z)^{2.98 \pm 0.55}\quad {\rm{or}}\quad n \approx 3  {\rm{.}}
\ee 
\noindent  Lubin and Sandage did not use the average, but assumed that the R-band galaxies had an evolution different from that of the I-band galaxies.  Due to their strong believe in the big-bang cosmology, they considered that these results indicated that the evolution had caused the galaxies to have greater surface brightness in the past, especially the R-band galaxies.  They found the experimental results to be inconsistent with the tired light model given by Eq.\,(A15), where $D_0 = D_{pl} =(c/H_0) \log (1+z)  {\rm{,}}$ because the experimentally determined exponent, about 3, is far from the exponent 1 in Eq.\,(A15).  We accept their conclusion about the tired light theory.

\indent  Lubin and Sandage did not know about plasma redshift, which predicts no coherent evolution (although each galaxy evolves).  For closer evaluation, we will in section 3 make more accurate comparison of the experiments and the plasma-redshift cosmology.  
\vspace{0mm}


\section{Translation of Tolman tests from big bang to plasma-redshift cosmology}

Lubin and Sandage give a procedure for translation of their results to different cosmologies.  In the present case it is, however, simpler to translate from big-bang cosmology to plasma redshift cosmology with help of the distances given by Eq.\,(A1), (A2), (A4) and (A6).

\indent  When we (as Lubin and Sandage) use the parameter $q_0 = 0.5$ in Eq.\,(A5), the distances in the big-bang cosmology are given by Eq.\,(A6), which for equal Hubble constants gives the same distances as Eq.\,(A4) when the cosmological parameters are $(\Omega_{\kappa},\,\Omega_m,\,\Omega_{\Lambda} ) = (0,\,1,\,0) {\rm{.}}~$ 

\indent   For small redshifts and for equal Hubble constants the distances determined by Eqs.\,(A1), (A2), (A4) and (A6) are all equal for the same $z$-values.  We need, therefore, to focus only on the increases in attenuation given by the different distances as redshift $z$ increases.  The ratios of the distances in the big-bang and plasma-redshift cosmologies vary with $z {\rm{;}}$ but the ratios are independent of the Hubble constant $H_0 {\rm{.}}~$ The variations of the attenuation factor $D(z)^2\, (1+z)^n$ in the denominator of Eq.\,(A13) and (A14) must for all cosmologies therefore equal the measured value.  We have therefore
\be
 D_{pl}^2 \, (1+z)^{n_{pl}}= D_{bb}^2\, (1+z)^{n_{bb}}~ ~~{\rm{or}}~~ ~\frac{{D_{pl}^2 }}{{D_{bb}^2}}=(1+z)^{n_{bb}-n_{pl}} 
\ee
\noindent  where $D_{pl}$ is plasma-redshift distance given by Eq.\,(A1), and $D_{bb}$ is the corresponding distance in the big-bang cosmology given by Eq.\,(A6).  More generally, the big-bang cosmologies depend on the cosmological parameters $(\Omega_{\kappa},\,\Omega_m,\,\Omega_{\Lambda} )$ used.  $n_{pl}$ and $n_{bb}$ are the exponents of $(1+z)$ for the plasma-redshift cosmology and the big-bang cosmology, respectively.  The measured values of $n_{bb}$ are determined by Lubin and Sandage [18\,-\,21], who used the big-bang cosmology and Eq.\,(A6) for estimating the distances.  The distant galaxies that they measured were in the clusters Cl~1324+3011 with $z=0.7565 {\rm{,}}$ Cl~1604+4304 with $z=0.8967 {\rm{,}}$ and in Cl~1604+4321 with $z=0.9243$ 
\vspace{2mm}

\indent  In Table 1, we list in columns 2, 3, and 4 the values $n_{bb}=n {\rm{,}}$ where the experimental $n$-values for the three clusters are listed as functions of $\eta$ in Tables 5, 6 and 7 by Lubin and Sandage [21].  For obtaining these measured values, Lubin and Sandage used Eq.\,(A6) for the distances, which are equal to those obtained by using Eq.\,(A4) with the big-bang parameters $(\Omega_{\kappa},\,\Omega_m,\,\Omega_{\Lambda} ) \approx (0.0,\,1.0,\,0.0) {\rm{.}}~$ Columns 5 and 6 of Table 1 are derived as explained in the following paragraph. 

\indent  In Fig.\,4 of reference [18], Lubin and Sandage show that for the R-band galaxies "the radius of the brightest galaxies changes more rapidly with $-M$ than $-0.2\, M {\rm{,}}$ which is the constant surface brightness condition".  When analyzing the data, we find that for $\Delta M_R(\eta)=-1 {\rm{,}}$ the $\log r {\rm{}}$-values increase beyond 0.2 by 0.22, 0.236, 0.22, and 0.27 for $\eta$ equal to 2, 1.7, 1.5, and 1.3, respectively.  This corresponds to the radii $r$ of the galaxies increasing by a factors of 1.66, 1.72, 1.66, and 1.86 when $\Delta M_R(\eta)=-1 {\rm{.}}~$ When Lubin and Sandage determined the radius of the distant R-band galaxies, they assumed big-bang expansion and divided therefore the observed radii, $r_{ex} {\rm{,}}$ by the factor $(1+z)=1.9243 {\rm{.}}~$ They therefore assumed that the R-band galaxies had smaller radii, $r {\rm{,}}$ than the actual value, $r_{ex} {\rm{,}}$ determined from a nonexpanding cosmology.  From these smaller radii, $r {\rm{,}}$ they concluded that the galaxies were dimmer than they actually are.  When they then measured the actual brightness they measured the right brightness, which was greater than that they expected from the small galaxies, because they had divided the observed radii $r_{ex}$ by the false expansion factor $(1+z) {\rm{.}}~$ The small radii, $r {\rm{,}}$ indicated to them that the galaxies were intrinsically dimmer than they actually were.  Had Lubin and Sandage realized that the radii were equal to the observed $r_{ex}{\rm{,}}$ they would have estimated the intrinsic brightness to be greater, and that the dimming caused by the high $z$-value was greater.  We would therefore have to increase the $n$-values.  For $\eta = 2 {\rm{,}}$ we get that $-\delta M_R =(\log 1.9243)/0.22 = 1.29$ and then $\delta n = (1.29/5)/\log 1.9243 = \mathbf{0.91} {\rm{.}}~$ For $\eta = 1.7 {\rm{,}}$ we get that $-\delta M_R =(\log 1.9243)/0.236 = 1.24$ and then $\delta n = (1.24/5)/\log 1.9243 = \mathbf{0.87} {\rm{.}}~$ For $\eta = 1.5 {\rm{,}}$ we get that $-\delta M_R =(\log 1.9243)/0.22 = 1.29$ and then $\delta n = (1.29/5)/\log 1.9243 = \mathbf{0.91} {\rm{.}}~$ For $\eta = 1.3 {\rm{,}}$ we get that $-\delta M_R =(\log 1.9243)/0.27 = 1.05$ and then $\delta n = (1.05/5)/\log 1.9243 = \mathbf{0.74} {\rm{.}}~$  The values are listed in the sixth column of Table 1.  In the fifth column we show what the exponent would be if there was no expansion for the R-band galaxies with the redshift $z = 0.9243 {\rm{.}}~$ These corrections apply only to the plasma redshift cosmology, which has no expansion.  


\begin{table}[h]
\centering
{\bf{Table 1}} \, \, The exponent $n_{bb}$ in the Tolman signal as a function of $\eta$ for $(\Omega_{\kappa},\,\Omega_m,\,\Omega_{\Lambda} ) \approx (0,\,1.0,\,0.0) {\rm{,}}$ 
\vspace{1mm}
\begin{tabular}{cccccc }
	\hline
~  ~ & $z = 0.7565$ &$z = 0.8967$ &$ z = 0.9243 $ &$ z = 0.9243 $ &$ z = 0.9243 $  \\
$\eta$ &$n_{bb}$ &$n_{bb}$ &$n_{bb}$ &$n_{bb} + \delta n_{bb}$ &$ \delta n_{bb} $  \\
	 \hline \hline
1.0 &  4.15 &  4.20 &  3.43 &    -    &    -      \\
1.3 &  3.81 &  3.93 &  3.15 &  (3.89) &  (0.74)   \\
1.5 &  3.81 &  3.91 &  3.11 &  (4.02) &  (0.91)   \\
1.7 &  3.48 &  3.50 &  2.76 &  (3.63) &  (0.87)   \\
2.0 &  3.25 &  3.29 &  2.48 &  (3.39) &  (0.91)   \\
\hline
\end{tabular}
\end{table}

\indent  Lubin and Sandage showed in their Fig.\,(1) of [18] that the values of $\eta$ in the range of $\eta=1.7~{\rm{to}} ~ \eta = 2.0$ are the most reliable.
\vspace{2mm}

\indent  Table 2 gives the experimental values Lubin and Sandage would have derived had they used the cosmology used by the supernovae researchers, who used the cosmological parameters $(\Omega_{\kappa},\,\Omega_m,\,\Omega_{\Lambda} ) \approx (0,\,0.3,\,0.7) {\rm{.}}~$  For small redshifts the distances are about the same as those in the other cosmologies, provided the Hubble constant is the same.  But for large redshifts, $(z=0.9243) {\rm{,}}$ the distances are 1.303 times larger than the distances given by Eq.\,(A6), which was used by Lubin and Sandage.  According to Eq.\,(5), the exponents in the Tolman signals, shown in column 2 to 4 of Table 2, are then significantly smaller than those in the corresponding columns of Table 1. The values for the R-band $( z = 0.9243) $ in the 5th column of each table are obtained by eliminating the expansion when estimating the brightness of these galaxies. 

\begin{table}[h]
\centering
{\bf{Table 2}} \, \, The exponent $n_{bb}$ in the Tolman signal as a function of $\eta$ for $(\Omega_{\kappa},\,\Omega_m,\,\Omega_{\Lambda} ) \approx (0,\,0.3,\,0.7) {\rm{,}}$ 

\vspace{1mm}

\begin{tabular}{ccccc }
	\hline
~$ $~ &~ $z = 0.7565$~ & ~$z = 0.8967$~ & ~$z = 0.9243$~&$ z = 0.9243 $  \\
$~\eta$~ & ~$n_{bb}$~ & ~$n_{bb}$~ & ~$n_{bb}$~& $n_{bb} + \delta n_{bb}$   \\
	 \hline \hline
1.0 &  3.31 &  3.38 &  2.62 &  -   \\
1.3 &  2.97 &  3.12 &  2.34 &  (3.08)   \\
1.5 &  2.97 &  3.10 &  2.30 &  (3.21)   \\
1.7 &  2.64 &  2.69 &  1.95 &  (2.82)   \\
2.0 &  2.41 &  2.48 &  1.67 &  (2.58)   \\
\hline
\end{tabular}
\end{table}

\indent  The exponents $n_{bb} {\rm{}}$ are smaller than those shown in Table 1 with $(\Omega_{\kappa},\,\Omega_m,\,\Omega_{\Lambda} ) \approx (0,\,1.0,\,0.0) {\rm{.}}~$  The values in the 5th column do not apply to the big-bang cosmology.  The deviations of $n_{bb}$ in columns 1 to 4 of Tables 1 and 2 from $n_{bb} = 4$ are blamed on the evolution.  Clearly the magnitude of the evolution depends on which big-bang model is correct.
\vspace{2mm}

\indent  In Table 3, we list for the galaxies in the three distant clusters the experimentally determined exponent $n_{pl}$ in plasma-redshift cosmology.  These values of $n_{pl}$ are obtained with help of Eq.\,(5) from the measured values $n_{bb}=n$ that are listed in Table 1 as a function of $\eta {\rm{.}}~$

\begin{table}[h]
\centering
{\bf{Table 3}} \, \, The exponent $n_{pl}$ in the Tolman signal as a function of $\eta$

\vspace{1mm}

\begin{tabular}{ccccc }
	\hline
~$ $~ &~ $z = 0.7565$~ & ~$z = 0.8967$~ & ~$z = 0.9243$~& ~$z_d = 0.9243 $~ \\
$~\eta$~ & ~$n_{pl}$~ & ~$n_{pl}$~ & $n_{pl} $ & $n_{pl} + \delta n_{pl}$  \\
	 \hline \hline
1.0 & 3.66  & 3.71  & (2.94)  & -  \\
1.3 & 3.32  & 3.44  & (2.66)  & 3.38  \\
1.5 & 3.32  & 3.42  & (2.62)  & 3.43  \\
1.7 & 2.99  & 3.01  & (2.27)  & 3.06  \\
2.0 & 2.76  & 2.80  & (1.99)  & 2.80  \\
\hline
\end{tabular}
\end{table}

\indent  For the R-band galaxies in the cluster Cl~1604+4321 with a redshift of $z=0.9243 {\rm{,}}$ we should in plasma-redshift cosmology shown in Table 3 use the 5th column and not the 4th.  This is because in plasma redshift cosmology, we do not divide the observed radius by $(1+z) {\rm{.}}~$ In the big-bang cosmology, the use of an expansion factor, $(1+z) {\rm{,}}$ leads to underestimation of the luminosity and the exponent in the Tolman signal, when evaluating the radius and the corresponding magnitude $M$.  It can be seen that columns 2, 3, and 5 are then remarkably similar.  This correction is significant as the comparison between column four and five illustrates.  We find thus that when evaluated in plasma-redshift cosmology, the experimental values for both the R-band and the I-band galaxies agree and are close to $n \approx 3$ in excellent agreement with Eq.\,(A13).

\indent  In plasma redshift cosmology, the distances are given by Eq.\,(A1), and the Hubble constant is given by Eq.\,(A3).  We see from Eq.\,(A3), that plasma redshift predicts intrinsic redshifts of galaxies, because the electron densities in the galactic coronas are greater than the average density in intergalactic space.  These intrinsic redshifts are often small.  In the Sun's corona the intrinsic redshift is about $\delta z=10^{-6} {\rm{.}}~$ In the corona of our Milky Way it is about $\delta z \approx 10^{-3} $ at latitudes $b\geq 20^o {\rm{.}}~$ In the corona of quasars the intrinsic redshift is often on the order of 1.  In case of older galaxies the redshift may be similar to that in the Milky Way, but in early type, it may be slightly greater.  These prediction match the observations [12\,-\,16].  Had we reduced the redshifts by the intrinsic redshift values, we would have increased the $n_{bb}$-values in Table 3 insignificantly.


\section{Evolution and plasma redshift}

In this section, we will see that the experimental evidences for an evolution that would reduce the exponent of (1+z) from 4 to about 3 in the denominator of Eq.\,(A14) are questionable, and we show how the different observed phenomena are well predicted by the plasma-redshift cosmology.

\indent  In the big-bang cosmology, we should have a general cosmological evolution of many phenomena.  The big-bang cosmologists expect therefore to see this evolution at work in the look-back time.  The over all formation rates of galaxies should have a beginning, reach a maximum, and then decrease.  The average of the universal present baryonic mass density should increase proportional to about $(1+z)^{3} {\rm{.}}~$ The energy density of the CMB photons should increase about proportional to $(1+z)^{4} {\rm{.}}~$ The average sizes of the galaxies should vary with time, etc.  Big-bang cosmologists often point to observations, which they believe indicate evolution.  We will analyze these observations and explain why they do not prove a coherent evolution.  We will show that the observed phenomena are consistent with non-expanding and quasi-static plasma-redshift cosmology.
\vspace{1mm}

\indent  {\bf{The big-bang cosmologists usually believe now that the cosmology is governed by the parameters $(\Omega_{\kappa},\,\Omega_m,\,\Omega_{\Lambda} ) \approx (0,\,0.3,\,0.7) {\rm{.}}~$}} The first four columns of Table 1 and 2 differ significantly.  If we accept the presently accepted big-bang cosmology, the evolution is much greater than that assumed by Lubin and Sandage.  
\vspace{1mm}

\indent  {\bf{The big-bang cosmologists did not know about plasma redshift}}.  They had therefore no explanation for the observed cosmological redshift except the Doppler effect.  They had no explanation for intrinsic redshifts.  They therefore denied that intrinsic redshifts existed.  Even the redshifts of quasars had to be explained as cosmological redshifts.  The intrinsic redshifts of galaxies in clusters are often enhanced by the intracluster plasma.  The assumed distribution of the velocities within the clusters is then too broad.  The big-bang cosmologists introduced then a "dark matter", which was invisible or undetectable except for the assumed gravitational effects.  They had no explanation for the cosmic microwave background (CMB) except an explosion of the universe, and they had no explanation for the soft X-ray background; nor could they explain the jets streaming from what they think are black holes.  They could not explain the hydrogen streaming from the center of our Milky Way.  They had no explanation for the heating of the intergalactic space; and they even could not explain the heating of the nearby solar corona.  In spite of these difficulties, they often believe very strongly that the big-bang model is correct.       
\vspace{1mm} 

\indent  {\bf{Plasma redshift on the other hand not only explains, but makes it necessary that we have the cosmological redshift, the CMB, and low-energy X-ray background}} [13].  Plasma redshift also makes clear that all stars and galaxies must have corona and therefore must have intrinsic redshifts.  It makes it clear that the hot plasma diffuses into intergalactic space and is heated by the redshift energy lost by the photons.  Hot large objects, especially when concentration of the high-Z element is low, must have large intrinsic redshifts.  A highly active star-bursting galaxy has a relatively high intrinsic redshifts.  These star bursting galaxies, even when close by, were assumed like the quasars to be far away.  The more than million K coronas of many objects are consequences of the plasma redshifts.
\vspace{1mm}

\indent  {\bf{The plasma redshift does not scatter the photons significantly}}, but the concurrent scattering on the plasma frequency in the intergalactic plasma causes a small scattering of about 200 pc at $z \approx 1.7  {\rm{;}}$ see Eq.\,(52) of [13].  This scattering is consistent with observations.  The Compton scattering removes the photons and contributes thereby to attenuations of light.
\vspace{1mm}

\indent  {\bf{Big-bang cosmologists also did not know that photons are weightless in the gravitational field}}.  A distant observer will even see the photons as if they were pushed away from a heavy objects.  All the experiments that have been assumed to prove the weight of photons have been incorrectly interpreted.  All these experiments were in the domain of classical physics [17].  Disregard for the uncertainty principle in quantum mechanics is the cause of the misinterpretation.  The weightlessness of photons (or gravitational repulsion of photons as seen by a distant observer) becomes clear when quantum mechanical effects are taken into account.  In the classical physics experiments, it was not possible to detect if the photons had a weight or were weightless.  However, the solar redshift experiments, which are in the domain of quantum mechanics, could and did confirm that the photons are weightless; see subsections 5.6 and section 6 of [13], and see the theoretical explanation in [17].  The photons are gravitationally redshifted in the Sun, but when the photons move from the Sun to the Earth, the photons lose their gravitational redshift.  The photons are being pushed outwards as seen by a distant observer [17].  Comparison of the observations and the predictions of the plasma redshift theory make this clear; see Fig.\,4 of [13].  This weightlessness of photons does not affect the bending of light, which is determined by Fermat's principle (50\,\% due to speed of light and 50\,\% due to curvature of space).  Nor does weightlessness affect the Shapiro effect.  Both of these effects are independent of the frequency of light.  The experimental evidence indicates that the weightlessness affects only photons and not for example the electromagnetic field of charged particles.  I call this "the modified theory of general relativity" [17].  This is a very significant modification of general relativity [13].
\vspace{1mm}

\indent  {\bf{Plasma redshift and the weightlessness of photons explain why there is no universal coherent evolution, although each star and each galaxy of course evolve}}.  Matter gradually concentrates and when the density and the pressure increase beyond a certain limit (close to the classical black hole limit), the matter can annihilate and transform into photons, which are repelled from the vortex of the objects close to the black hole limit.  The photons transform through pair production (electron-positron and proton-antiproton pairs, and heavier particle pairs) thereby renewing matter, including hydrogen; see section 6 of [13].  Matter enhances the pair production.  Once the matter concentrates at a certain spot on the jet stream of photons, that matter spot has a tendency to grow.  These processes are consistent with observations and physics as we know it and can go on forever.  In plasma redshift cosmology, the universe renews itself and lasts forever. 
\vspace{1mm}

\indent  In section 7 of their paper [21], Lubin and Sandage state:
 {\bf{"The galaxies in the three clusters studied here have fainter absolute magnitudes and smaller radii than the average local clusters"}}.  They see this as an evolutionary trend.

\indent  {\bf{The faintness is caused partially by the longer distances in the plasma-redshift cosmology than in the big-bang cosmology used by Lubin and Sandage, and partially by the factor}} ${\mathbf{ (1+z)^{0.5} }} {\rm{.}}~$ The distances are determined by Eq.\,(A1) and Eq.\,(A6), respectively.  For small redshifts and equal Hubble constants, the attenuations of emitted light and sizes of the galaxies are identical for these two cosmologies.  For both cosmologies the distances are $D \approx (c/H_0)\,z$ and $(1+z) \approx 1 {\rm{.}}~$ But for high-z galaxies the distances differ, and the factor $(1+z)> 1 {\rm{.}}~$ For example, for the galaxies in the cluster Cl 1604+3011 with $z=0.8967$ the distance in plasma redshift cosmology is 1.169 times that determined by Eq.\,(A6) used by Lubin and Sandage.  In addition, the plasma redshift cosmology has a factor $(1+z)^{1.5} {\rm{}}$ in the luminosity distance, see Eq.\,(A9), while the big-bang cosmology has a factor of only $(1+z) {\rm{;}}$  see Eq.\,(A10).  {\bf{The total intensity in plasma-redshift cosmology is therefore}} $(D_{bb}/D_{pl})^2 /(1+z) = 0.73/1.8967 ={\mathbf{0.386}}$ times that in the big-bang cosmology used by Lubin and Sandage.  This dimming corresponds to a magnitude change of ${\mathbf{\Delta M = 1.03}}=- 2.5\, \log 0.386$ relative to that expected from their assumed cosmology.  This dimming matches 1.04 observed by Lubin and Sandage.  (The 1.04-value is derived from their Table 8 of [21] for plasma redshift.)

\indent  {\bf{The radii of the galaxies appear smaller}}, because the distances in plasma-redshift cosmology given by Eq.\,(A1) are 16.9\,\% greater than the distances given by Eq.\,(A6) used by Lubin and Sandage; and because the big-bang cosmologists expect the apparent radius of the galaxy to increase by the factor $(1+z) {\rm{.}}~$ {\bf{The estimated radii of the galaxies in this cluster are then smaller than those expected by a factor of}} $1/(1.169\cdot 1.8967)={\mathbf{0.451}} {\rm{.}}~$ We find that this reduction is consistent with that observed by Lubin and Sandage; see Figs.\,(1) and (2) of [21].

\indent  {\bf{The apparent surface brightness $i_{\langle SB \rangle} {\rm{,}}$ according to Eqs.\,(A13) and (A14), will be}} $0.73\cdot 1.8967 = {\mathbf{ 1.39}}$ times greater than that expected by Lubin and Sandage from Eq.\,(A1).  These apparent changes in the luminosity and surface brightness are by Lubin and Sandage referred to as luminosity evolution, although no such universal evolution takes place according to plasma-redshift cosmology, which is consistent with observations.
\vspace{1mm}

\indent  {\bf{Barden et al.}} [29] {\bf{did similar studies as Lubin and Sandage}}.  Had we, as Barden et al. did [29], in the flat big-bang cosmology used the parameters: $(\Omega_m ,\, \Omega_{\Lambda}) = ( 0.3, \,0.7 ) {\rm{,}}$ we would for small redshifts derive similar distances, surface intensities, and total magnitudes as those in the plasma-redshift cosmology.  But for $z = 1 {\rm{,}}$ the distance in the big-bang cosmology would for be $(c/H_0)\, 0.7714 ~{\rm{Mpc}}$ or 1.113 times larger than the distance $(c/H_0)\, 0.6931~{\rm{Mpc}}$ in the plasma-redshift cosmology.  The expansion parameter would be $(1 + z) = 2 {\rm{.}}~$ The big-bang cosmologists would then see the radius of the distant galaxies shrink by a factor of $1.113/2 ={\mathbf{ 0.556}} {\rm{}}$ relative to similar galaxies close by.  {\it{This is consistent with that observed by Barden}} [29], see their Fig.\,7.

\indent  {\bf{The distant galaxies observed by Barden et al.}} [29] {\bf{should appear fainter by a factor of}} $1.113^2 /2 ={\mathbf{ 0.62 }}{\rm{,}}$ which corresponds to magnitude change of $\Delta M = 0.52 {\rm{.}}~$ The brightness intensity would increase by a factor of $1.113^2 \cdot 2 ={\mathbf{ 2.48}} {\rm{,}}$ which corresponds to a surface magnitude change of $\Delta M ={\mathbf{-0.99}} {\rm{,}}$ {\it{which is consistent with the $dM/dz = -0.99$ observed by Barden et al.}} [29]; see end of their section 4.1.  These numbers explain why Barden et al. think there is an evolution when in fact there is no coherent evolution according to the plasma redshift theory.  Like Lubin and Sandage, Barden et al. [29] concluded that the evolution was significant.  But {\it{plasma redshift assumes no coherent evolution and explains the observations as being consistent with these predictions of the plasma-redshift cosmology.}}
\vspace{1mm}

\indent  In section 1.4.1 of their paper [21], Lubin and Sandage refer to the "{\bf{spectacular confirmation of the time dilation effect}}" by Goldhaber et al, [5].

\indent  Lubin and Sandage are certainly right that these experiments are usually considered to confirm time dilation.  However, as Brynjolfsson in [13]\,and\,[16] has shown, the experiments clearly show that {\bf{the interpretation by Goldhaber et al.}} [5] {\bf{is wrong}}.  In their analysis Goldhaber et al. did not take the Malmquist bias into account.  It becomes clear that if the Malmquist bias is taken properly into account there is no time dilation.  The experimental analysis by Guy et al.~[8] of the changes in magnitude versus light-curve width underscores that the relationship in Figure 3 of [5] is due to Malmquist bias, and {\bf{that there is no time dilation}}; see Brynjolfsson [16].  The supernovae researchers strongly believed in the big bang and its time dilation.  They therefore divided the light curve width by the time dilation factor, which reduces the absolute magnitude of the distant supernovae explosions relative to those nearby.  Thus their determination of both the absolute magnitude and the cosmological parameters is wrong.
\vspace{1mm}

\indent  In section 1.4.2 of their paper [21], Lubin and Sandage refer to reports, which reported that "{\bf{observations of the Boltzmann temperature of interstellar molecules in the spectra of high-redshift galaxies has now apparently been measured in a difficult experiment}}", which indicated a higher temperature of the cosmic microwave background (CMB) in the past. 

\indent  I have failed to find evidence that the temperature of interstellar molecules is determined by CMB.  The analyses by Spitzer [30] indicate that the temperature of interstellar matter including molecules is not determined by CMB, but by collision with surrounding atoms molecules and dust particles, and by higher energy photons and X-ray radiation.  The interstellar medium has a very broad and a continuous spectrum of temperatures within both the H\,I and the H\,II regions.  The tremendous difference in temperature between the hot regions ($T\approx 10^6~{\rm{K}}$) and the colder regions of space ($T\leq 10^4~{\rm{K)}}$ is presently explainable only with help of the plasma redshift.

\indent  The formation of the huge hot plasma bubbles in interstellar and intergalactic space (as well as in the transition zone to the solar corona) [13] comes about because the plasma redshift is proportional to the electron density, $N_e {\rm{,}}$ while the cooling processes are about proportional to $N_e^2 {\rm{.}}~$ The high temperatures are mainly limited by the heat conductivity.  The heat conduction from the hotter volumes and the X-ray heating counteract the lowest temperatures in the H\,II regions.  Plasma-redshift heating, together with the heating by high-energy photons and X rays are the major component of the heating of interstellar and intergalactic medium.  The hot intergalactic plasma absorbs CMB; but on the average, this absorption is balanced by the corresponding CMB emission.  In the colder H\,I regions the temperature is controlled mainly by the collisions.
\vspace{1mm}

\indent  {\bf{The blackbody spectrum of the CMB is often sited as a proof of the big bang}}.  Peebles [1] points out that the absorption of CMB in intergalactic space would deform the spectrum.

\indent  {\bf{However, in plasma redshift cosmology, equal amount of CMB is absorbed as is emitted from the intergalactic plasma}}, see section 5.10 of [13].  No deformation of the spectrum is therefore observed.  The plasma redshift dominates by far all other absorption and emission processes in this frequency range.  This accounts for the nice blackbody spectrum.  The average electron density  $N_e\approx 0.0002~{\rm{cm}}^{-3} {\rm{,}}$ and the average particle temperature $T_e \approx 2.7\cdot 10^6~{\rm{K}}$ of the intergalactic plasma produces the CMB radiation with the temperature of $T_{CMB}\approx 2.73$ in a volume with a radius equal to one Hubble length, about 5000 Mpc [13].  The big-bang cosmologists surmise incorrectly that the radiation temperature of CMB emitted from a plasma at the time of decoupling is proportional to the particle temperature at that time.  This is based on faulty physics.  The temperature $T_{CMB} = (0.1049\,N_e T_e)^{1/4}$ of the blackbody spectrum of the CMB emitted by sparse plasma is much lower than the particle temperature; see sections 5.10 and 5.11 and Appendix C of [13].  The errors made by the big-bang cosmologists are caused mainly by the fact that they did not know the plasma redshift.
\vspace{1mm}

\indent  {\bf{Plasma redshift should not be confused with the static model by Zwicky}}, a conjecture without any support from physics, while plasma redshift is based on hard core physics and follows from conventional well-established axioms of physics.  For example, plasma-redshift demands concurrent Compton effect, which Zwicky's model did not contain.  In the hot intergalactic plasma, the Compton cross section is exactly twice that of the plasma redshift cross-section.  Furthermore, plasma-redshift supplies directly and indirectly most of the needed heating of the intergalactic plasma.  Plasma redshift gives a natural explanation of the blackbody CMB radiation in a blackbody cavity with radius equal to the Hubble length.  Plasma redshift explains directly the Olber's paradox, because all light is attenuated at least by redshift factor of $e^{-1}=0.37 $ and by the Compton scattering factor $e^{-2}=0.14 $ for a total of $e^{-3}=0.05 $ in one Hubble length.  In addition the intensity of course decreases inversely proportional to the distance squared, $D_{pl}^{-2} {\rm{.}}~$ Other processes such as scattering and absorption by bound electrons in atoms are usually significant.

\indent  Plasma redshift should not be confused with the steady state model conjectured by Bondi and Gold [32] or that by Hoyle [33], which as Lubin and Sandage correctly characterized as a different kind of an expansion models [21].

\vspace{0mm}


\section{Conclusions}

We mainly used the analyses of the experimental data by Sandage and Lubin [18 - 21] of the variation in surface brightness of galaxies with the redshift to show that there is no cosmic expansion and that the experimental data give a good fit to the plasma-redshift cosmology.

\indent  In section 3, we showed how to translate the different cosmologies for more accurate comparison with the experimental data.  Lubin and Sandage replaced the factor, $(1+z)^4 {\rm{,}}$ in the denominator of Eq.\,(A14) by $(1+z)^n$ and determined $n$ from the experiments.  In their Fig.\,1 of [18], they showed that the brightness is best determined when the parameter $\eta$ in Eq.\,1 is between 1.7 and 2.

\indent  The exponent derived from their experimental data for their big-bang $(\Omega_{\kappa},\,\Omega_m,\,\Omega_{\Lambda} ) \approx (0,\,1.0,\,0) {\rm{}}$-cosmology are listed in columns 2, 3, and 4 of Table 1. For the I-band galaxies ($z=0.7565$ and $z=0.8967$) the average for $\eta =1.7$ and $\eta =2$ is $n =\mathbf {3.38} {\rm{;}}$ and the corresponding average for the R-band galaxies ($z=0.9243$) is $n=\mathbf{2.62} {\rm{.}}~$ The over all average is $n_{av}=\mathbf{3.00} {\rm{.}}~$ The corresponding evolution in the big-bang cosmology used by Lubin and Sandage is then $4-n_{av}=\mathbf{1.00} {\rm{.}}~$ 

\indent  For the now preferred $(\Omega_{\kappa},\,\Omega_m,\,\Omega_{\Lambda} ) \approx (0,\,0.3,\,0.7) {\rm{}}$-cosmology the exponent $n$ is shown in columns 2, 3, and 4 of Table 2.  The corresponding averages of $n$ for $\eta =1.7$ and $\eta =2$ for the I-band galaxies is $n=\mathbf{2.56} {\rm{,}}$ and for R band $n=\mathbf{1.81} {\rm{.}}~$ The over all average is $n_{av}\mathbf{=2.18} {\rm{.}}~$ The evolution in the now conventional big-bang cosmology is then about $4-n_{av}=\mathbf{1.82} {\rm{.}}~$

\indent  In the plasma-redshift cosmology shown in columns 2, 3, and 5 of Table 3, the corresponding averages for $\eta =1.7$ and $\eta =2$  are for I-band galaxies $n=\mathbf{2.89}$ and for R-band galaxies $n=\mathbf{2.93} {\rm{.}}~$  The over all average is $n_{av}=\mathbf{2.91} {\rm{.}}~$ The corresponding evolution from the expected $n=3$ in Table 3 is then $3-n_{av}=\mathbf{0.09} {\rm{.}}~$ This shows that there is about no evolution, and that the exponent is  $n \approx 3 {\rm{,}}$ as predicted by the plasma-redshift cosmology. 

\indent  In the big-bang cosmologies the values of $n$ for the I-band and R-band galaxies differ significantly.  But in the plasma redshift cosmology, the values of $n=\mathbf{2.89}$ for the I band and $n=\mathbf{2.93} {\rm{}}$ for the R band are practically equal.  This is an independent confirmation of that there is no expansion, because as shown in section 3, the difference between I band and Rband in the big-bang cosmologies is rooted in the assumed dilation factor.

\indent  We see also that the values of $n$ derived in the two big-bang cosmologies differ significantly.  The evolution $4-n_{av} \approx 1.82$ in the now preferred big-bang cosmology is much greater than $4-n_{av} \approx 1.00$ in the cosmology used by Lubin and Sandage.  In case of the now preferred cosmology, the evolutions surmised by Lubin and Sandage must be increased significantly. 

\indent  In section 4, we considered the many indicators of evolution mentioned by Lubin and Sandage.  We found that the observations are better explained by the plasma-redshift cosmology, which predicts no expansion and no coherent evolution.  Lubin and Sandage referred to the experiments by Goldhaber et al. [5], which are widely surmised to have confirmed the expansion.  We have shown [13]\,and\,[16] that when we take the Malmquist bias properly into account, these experiments by Goldhaber et al. actually show that there is no expansion.  This has the consequence that the absolute magnitude of the supernova is incorrect.  Thus, the determination of the cosmological big-bang parameters, $(\Omega_m,\,\Omega_{\Lambda}) = (0.3,\,0.7) {\rm{,}}$ by the supernovae researchers is incorrect.  We showed also that observations of the CMB and X rays are well predicted by the plasma-redshift cosmology.  The present comparison of experiments and theory underscores that there is no expansion.

\indent  In section 4, we considered also Lubin and Sandage statement that: {\it{"The galaxies in the three clusters studied here have fainter absolute magnitudes and smaller radii than the average local clusters".}}  These interpretations of the observations are due to the incorrect distances and the false expansion in the big-bang cosmology.  In plasma-redshift cosmology the magnitude and the dimensions of these galaxies appear about equal to those of nearby galaxies.  There is no need for expansion or coherent evolution.  This again indicates that any expansion cosmology is incorrect.
\vspace{5mm}


\renewcommand{\theequation}{A\arabic{equation}}
\setcounter{equation}{0}
\section*{Appendix A \, \, Distances in the different cosmologies}


\subsection*{A1 \, \, Distance-redshift relations in the different cosmologies}

In the {\bf{plasma redshift cosmology}}, the distance-redshift relation (see Eq.\,(50) in reference [13]) is rather simple and has no adjustable parameters.  It follows strictly from the average electron density in intergalactic space.  It is given by
\be
 D_{pl} =\frac{{c}}{{H_0}} \cdot \ln (1+z){\rm{,}}
\ee
where $D_{pl}$ is in megaparsec (Mpc), $c$ the velocity of light in ${\rm{km\,s}}^{-1}{\rm{,}}$ and $z$ is the redshift along the line from the source to the observer.  $H_0$ is the Hubble constant in ${\rm{km\,s}}^{-1}{\rm{Mpc}}^{-1}{\rm{.}}~$ For small redshifts, we have that $\ln\,(1+z)\approx z $, and the distances are equal to 
\be
D_{pl} = \frac{{c}}{{H_0}}  \cdot z,
\ee
\noindent  as original discovered by Hubble.  In plasma redshift cosmology, the Hubble constant is proportional to the average electron density along the line from the observer to the object (see Eq.\,(49) of [13]),
\be
H_0 = 3.076\cdot 10^5 \cdot (N_e)_{av}~~{\rm{km\,s}}^{-1}{\rm{Mpc}}^{-1}{\rm{,}}
\ee
where $(N_e)_{av} ~{\rm{cm}}^{-3}$ is the average electron density along $D_{pl}{\rm{}}$ in cm [13].
   

\begin{figure}[t]
\centering
\includegraphics[scale=.5]{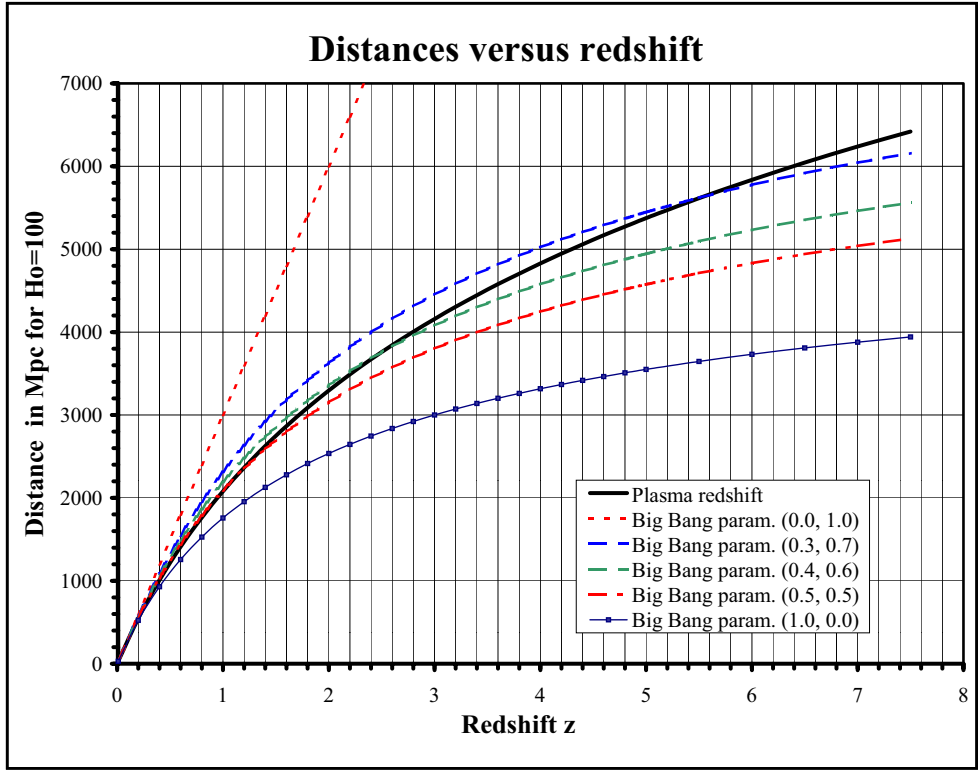}
\caption{The ordinate gives the distance in Mpc for a Hubble constant $H_0 = 100\,{\rm{km\,s}}^{-1}\,{\rm{Mpc}}^{-1}$ versus the redshift, $z$ from 0 to 8 on the abscissa.  The black heavy solid curve gives the distance in a cosmology based on the plasma redshift.  The remaining 5 curves give the distances based on the flat big-bang cosmology when the cosmological parameters are $(\Omega_m,\,\Omega_{\Lambda}) = (0.0, \, 1.0), \,(0.3, \, 0.7), \, (0.4, \, 0.6), \, (0.5, \, 0.5), \,{\rm{and}}\, (1.0,\,0.0), $ respectively.  It can be seen that we can adjust the cosmological parameters in the big-bang cosmology to give a fairly good fit to the solid curve valid for the plasma-redshift theory.  A better fit is obtained if we also vary the parameters with the redshift, z. }
\end{figure}

\noindent  The {\bf{contemporary big-bang cosmology}} uses three adjustable parameters $\Omega_k ,\, \Omega_m ,\,{\rm{and}}\,\Omega_{\Lambda} {\rm{}}$ for estimating the comoving distances $D_{bb}$ given by (see [1] or Eq.(B1) of [11])
\be
D_{bb} = \frac{{c}}{{H_0}}\,\frac{{1}}{{ |\Omega_k|^{1/2} }}\,{\rm{sinn}}\left[ \int_0^z \frac{{ |\Omega_k|^{1/2} \,dz'}}{{\,\left[ (1+z')^2(1+{\Omega}_m \,z') - z'(2+z') {\Omega_{\Lambda}} \right]^{1/2} \,}}\right]{\rm{.}}
\ee
\noindent  Usually, the curvature parameter $\Omega_k $ is set equal to $\Omega_k = 1- \Omega_m - \Omega_{\Lambda}{\rm{,}}$ where $\Omega_m = 8\pi \rho /(3H_0^2 )$ is the mass density parameter, and where $\rho$ in ${\rm{g\,cm}}^{-3}$ is an adjustable parameter, which includes both the dark matter and baryonic matter densities.  The dark energy parameter $\Omega_{\Lambda} = \Lambda/(3H_0^2)$ is an expansions parameter, and $\Lambda$ the cosmological constant (an adjustable parameter).  Often used values for these parameters in flat space (that is for $\Omega_k = 0 $) are $(\Omega_m , \, \Omega_{\Lambda}) = (0.3, \, 0.7).~$  The function sinn\,(x) is equal to $\sinh (x)$ for $\Omega_k > 0$, equal to $x$ for $\Omega_k = 0 ,$ and equal to $\sin(x)$ for $\Omega_k < 0 {\rm{.}}~$ These parameters are usually adjusted to fit the observations.

\indent  In the limit of $\Omega_k = \Omega_m = \Omega_{\Lambda} = 0 {\rm{,}}$ Eq.\,(A4) becomes identical to Eq.\,(A1), which is valid in the plasma-redshift cosmology.  In the limit of $\Omega_k = \Omega_m = 0,~ {\rm{and}}~ \Omega_{\Lambda} = 1 {\rm{,}}$ Eq.\,(A4) becomes identical to Eq.\,(A2); and for small $z$-values, both Eqs.\,(A4) and (A1) are identical to (A2).


\begin{figure}[t]
\centering
\includegraphics[scale=.5]{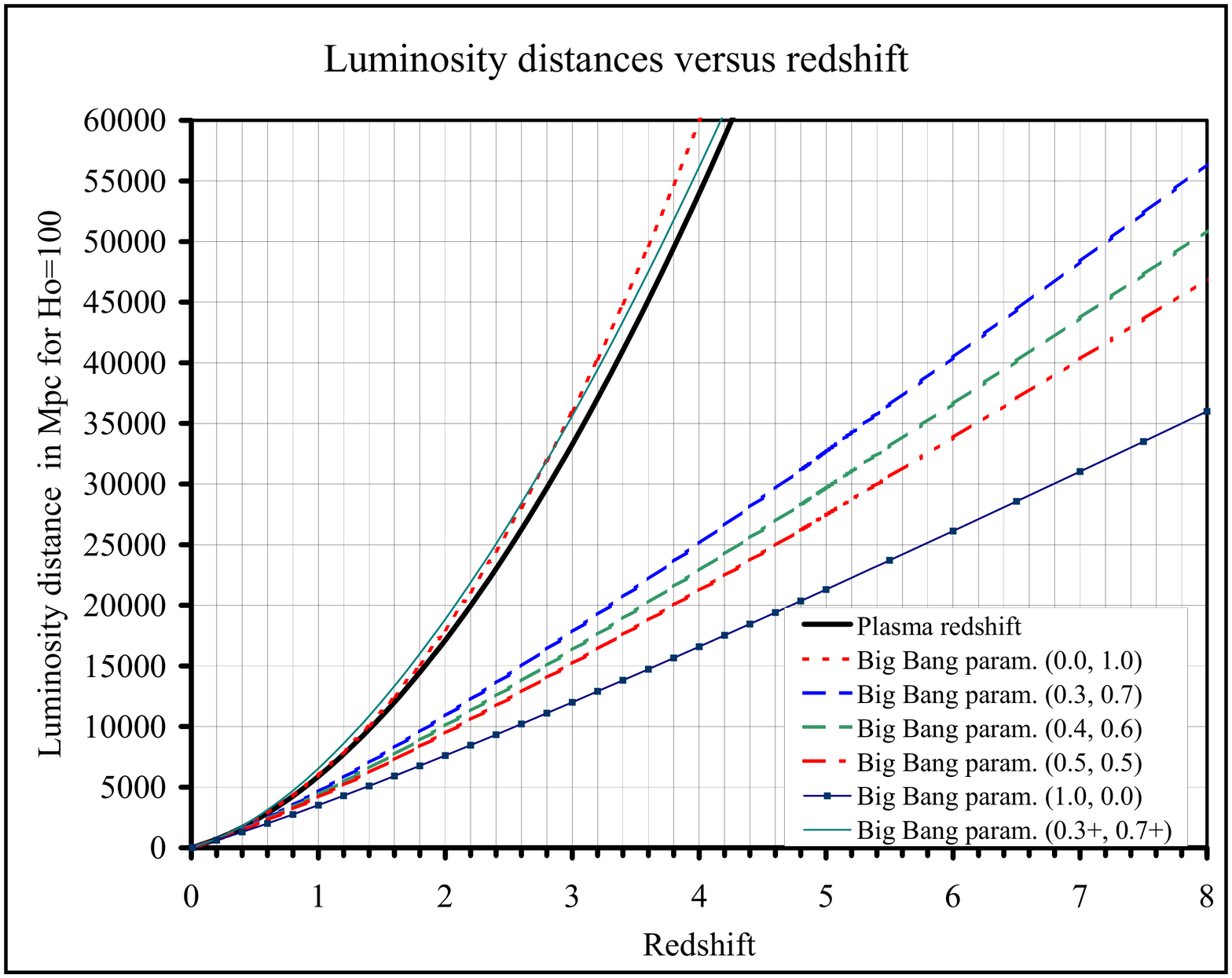}
\caption{The ordinate gives the luminosity distance in Mpc for a Hubble constant $H_0 = 100\,{\rm{km\,s}}^{-1}\,{\rm{Mpc}}^{-1}$ versus the redshift from $z = 0$ to $z = 8$ on the abscissa.  The heavy black solid curve gives the luminosity distance in a cosmology based on the plasma redshift, while the next 5 curves give the distances based on the flat big-bang cosmology when the cosmological parameters $(\Omega_m,\,\Omega_{\Lambda}) = (0.0, \, 1.0), \,(0.3, \, 0.7), \,  (0.4, \, 0.6), \, (0.5, \, 0.5), \,{\rm{and}}\, (1.0,\,0.0){\rm{,}} $ respectively.  The thin solid curve with the big-bang parameters (0.3+,\,0.7+) next to the heavy black solid curve is obtained by multiplying the curve with parameters (0.3,\,0.7) by $(1+z)^{0.5}{\rm{.}}~$ It can been seen that the curves with the big-bang parameters (0.0,\, 1.0) and (0.3+,\, 0.7+) are close to the heavy black solid curve valid for the plasma-redshift theory.  A better fit is obtained if we also let the parameter $\Omega_{\Lambda}= 0.7+$ increase slightly with the redshift, z. }
\end{figure}

\indent  The big-bang cosmologists sometimes use the {\bf{Mattig's equation}}; see [34] or Eq.\,(30) of [35] or Eq.\,(1) of [21]
\be
D_{mat} = \frac{{c}}{{ H_0\, q_0^2\, (1+z) }}\,\left[ z\,q_0 +(q_0 -1)\, \{ -1 + \sqrt{2q_0\,z + 1\,}\,  \}\right]
\ee  
\noindent  Sandage and Lubin [18\,to\,21]  set $q_0 = 0.5 $ in Eq.\,(A5) and get then
\be
D_{mat} = D_{bb} = \frac{{ 2c\, (\sqrt{1+z\,}-1\,) }}{{ H_0\, (1+z)^{1/2}\, }} = \frac{{ c }}{{ H_0 }}\, 2 \left[ 1 - \frac{{1}}{{ \sqrt{1+z\,} }} \,\right] 
\ee
\noindent  In the limit of $\Omega_k = \Omega_{\Lambda} = 0 ~{\rm{and}}~\Omega_m = 1 {\rm{,}}$ Eq.\,(A4) becomes identical to Eq.\,(A6).
 
\indent Fig.\,1 illustrates how the distances in flat space vary with the redshift.  It indicates that the parameters, $\Omega_m,\,\Omega_{\Lambda},~{\rm{and}}~\Omega_k $ can be adjusted to give a fairly good fit to the plasma-redshift theory.  Besides adjusting these cosmological parameters, we can also adjust the Hubble constant.

\indent  The supernovae researchers usually adjusted the parameters to be $(\Omega_m,\,\Omega_{\Lambda}) \approx (0.3,\,0.7)$ and a flat space, $\Omega_k  = 0 {\rm{.}}~$ They also use a slightly higher Hubble constant than that I derive from plasma-redshift theory.  Their fit of the distances are then about equal to those derived from the plasma redshift, see the heavy black curve in Fig.\,1.
\vspace{3mm}


\subsection*{A2 \, \, The magnitude-redshift relation and the luminosity distances}

\indent  {\bf{In the plasma-redshift cosmology}}, the magnitude-redshift relation is given by (see section 5.8 in reference [13]) 
\be
m - M = 5 \log \left( D_{pl} \right)  + 2.5 \log \left( {1 + z} \right) + 5 \log \left( {1 + z} \right) + 25 \,{\rm{.}} 
\ee
\noindent  On the left side $m = m_{obs} - 1.086 a {\rm{,}}$ where $m_{obs}$ is the observed magnitude and $a{\rm{}}$ the extinction, which should include the extinction caused by Compton and Rayleigh scatterings on bound electrons in atoms.  $M$ is the absolute magnitude.  The first term on the right side is due to the distance $D_{pl}$ in Mpc, as given by Eq.\,(A1).  The second term, $2.5 \log (1+z){\rm{,}}$ is due to reduction in photon energy by plasma redshift, and the third term, $5 \log (1+z){\rm{,}}$ is due to removal of photons through Compton scattering on the free electrons in the intergalactic plasma.  The last term corrects for the units used. 

\indent  {\bf{In big-bang cosmology}}, the magnitude-redshift relation is given by  
\be
m - M = 5 \log \left( D_{bb} \right) + 2.5 \log \left( {1 + z} \right) + 2.5 \log \left( {1 + z} \right) + 25 {\rm{,}}
\ee
\noindent  where the first term on the right side is due to the distance $D_{bb} {\rm{.}}~$ The second term is due to the cosmological redshift and the third term is due to the cosmological time dilation.
\vspace{3mm}  

\noindent  {\bf{The luminosity distance $D_{lu}$}} shown in Fig.\,2 is often used by the big-bang cosmologists.  It is the distance that gives the same dimming of the stars (or galaxies) as the three first terms on the right side of Eq.\,(A7) and (A8), respectively.  In plasma-redshift cosmology we have therefore that the luminosity distance $D_{lu}$ is 
\be
D_{lu} = D_{pl} \, (1+z)^{1.5} \, {\rm{.}}
\ee
\noindent  In big-bang cosmology we analogously that the luminosity distance $D_{lu}$ is
\be
D_{lu} = D_{bb} \, (1+z) \, {\rm{.}}
\ee
\noindent  We can then replace the 3 first terms on the right side of Eq.\,(A7) and (A8) by one term, $5 \log D_{lu} {\rm{.}}$

\indent  Fig.\,2 illustrates how the luminosity-distances for the different cosmologies varies with the redshift.  According to Eqs.\,(A7) and (A8) the total light intensity decreases then in plasma-redshift cosmology as
\be
\quad I\propto\frac{{I_0}}{{D_{lu}^2}}=\frac{{I_0}}{{D_{pl}^2 \, (1+z)^{3}}} \quad  \quad \quad \quad \quad \quad ~
\ee
while in big-bang cosmology the total intensity decreases as
\be
\quad \quad ~~~I \propto \frac{{I_0 }}{{D_{lu}^2}}= \frac{{I_0}}{{D_{bb}^2 \, (1+z)^2}} \quad \quad \quad  \quad \quad \quad ~~
\ee 

\noindent  These total intensities determine the magnitude-redshift relation in Eqs.\,(A7) and (A8).
\vspace{3mm}


\subsection*{A3 \, \, The surface brightness distances in the different cosmologies}
   

\begin{figure}[t]
\centering
\includegraphics[scale=.5]{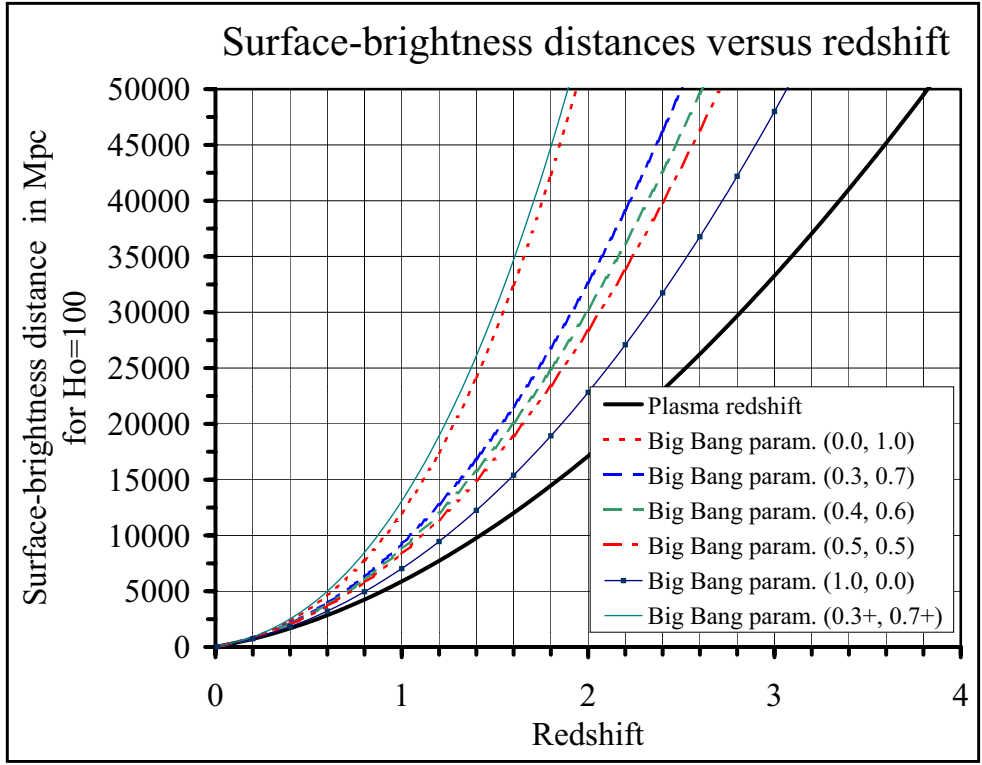}
\caption{The ordinate gives the surface brightness distance in Mpc for a Hubble constant $H_0 = 100\,{\rm{km\,s}}^{-1}\,{\rm{Mpc}}^{-1}$ versus the redshift from $z = 0$ to $z = 8$ on the abscissa.  The heavy black solid curve gives the surface brightness distance in a cosmology based on the plasma redshift, while the next 5 curves above it give the surface brightness distances based on the flat big-bang cosmology when the cosmological parameters $(\Omega_m,\,\Omega_{\Lambda}) = (0.0, \, 1.0), \,(0.3, \, 0.7), \,  (0.4, \, 0.6), \, (0.5, \, 0.5), \,{\rm{and}}\, (1.0,\,0.0), $ respectively.  The thin solid curve above the others has the big-bang parameters (0.3+,\,0.7+) which are obtained by multiplying the curve with parameters (0.3,\,0.7) by $(1+z)^{0.5}{\rm{.}}~$ It can been seen that the curve with the big-bang parameters (0.0,\, 1.0), which corresponds to Mattig's equation with parameter $q_0$ used by Lubin and Sandage is the closest curve to the plasma-redshift curve.}
\end{figure}

\noindent  {\bf{The surface brightness}} is the light intensity per unit area.  When integrated over the entire area of the galaxy, it gives the total light intensity that determines the magnitude.  In the big-bang cosmology the hypothetical expansion causes the area of the galaxy to appear as if expanding during light's travel.  This reduces the surface brightness by a factor of $(1+z)^{-2} {\rm{.}}~$  In plasma-redshift cosmology there is no expansion, and therefore no such additional factors needed.  If we can measure the light intensity per square area of the galaxies, we should be able to find out if this predicted expansion in the big-bang cosmology is real.  In plasma-redshift cosmology the surface brightness $i_{\langle SB \rangle}$ is given by
\be
i_{\langle SB \rangle}\propto \frac{{(i_{\langle SB \rangle})_0 }}{{D_{sb}^2 }} = \frac{{(i_{\langle SB \rangle})_0}}{{D_{lu}^2}}=\frac{{(i_{\langle SB \rangle})_0 }}{{D_{pl}^2 \, (1+z)^{3}}}
\ee  
where $(i_{\langle SB \rangle})_0 $ is the surface brightness at the time of emission, and $D_{sb}= D_{pl} \, (1+z)^{1.5}$ is the surface-brightness distance.  In big-bang cosmology the surface brightness is given by
\be
i_{\langle SB \rangle} \propto \frac{{(i_{\langle SB \rangle})_0}}{{D_{sb}^2 }} =\frac{{(i_{\langle SB \rangle})_0  }}{{D_{lu}^2 (1+z)^2}}= \frac{{(i_{\langle SB \rangle})_0}}{{D_{bb}^2 \, (1+z)^4}} 
\ee 
\noindent  where $(i_{\langle SB \rangle})_0 $ is the surface brightness at the time of emission, and $D_{sb}= D_{bb} \, (1+z)^{2}$ is the surface-brightness distance.  Eq.\,(A13) is similar to Eq.\,(A11), because in plasma-redshift cosmology there is no expansion.  On the other hand, in big-bang cosmology the expansion increases the denominator in Eq.\,(A14) by a factor of $(1+z)^2$ over the denominator of Eq.\,(A12).

\indent  Fig.\,(3) illustrates how the surface brightness distances in the different cosmologies vary with the redshift. 

\indent  If we integrate the surface brightness to derive the total intensity, we get a good test for distinguishing the correctness of the two cosmologies.

\indent  In the denominator on the right side of Eq.\,(A13) one $(1+z)$-factor of $D_{pl}^2 \, (1+z)^3$ is due to the decrease in energy flux caused by the redshift, and the remaining factors $(1+z)^2$ are due to the Compton scattering with a cross section exactly twice that of the plasma redshift [13].  There is no time dilation in the plasma redshift cosmology.

\indent  In the denominator on the right side Eq.\,(A14), we have that one $(1+z)$-factor in $D_{bb}^2 \, (1+z)^4$ is due to the decrease in energy caused by the redshift; the second $(1+z)$-factor is caused by the time dilation, which results in longer times between the arriving photons at the position of the observer.  The two remaining factors $(1+z)^2$ are due to the expansion, which increases the area of the galaxy (object) as seen by the observer.

\indent  Lubin and Sandage [21] compared Eq.\,(A14) with the tired light theory proposed first by Zwicky in 1929 [31].  In the tired light model we have 
\be
i_{\langle SB \rangle}  \propto \frac {{1}}{{D_{tl}^2 \,(1+z) }}\, {\rm{,}}    
\ee   
\noindent with only one factor $(1+z)$ in the denominator instead of three and four such factors in Eq.\,(A13) and (A14), respectively.  As Lubin and Sandage pointed out, the tired light theory is excluded by the experiments, which show that $n\approx 3 {\rm{.}}~$


\end{document}